\documentclass[aps,prl,twocolumn,superscriptaddress,showpacs,10pt]{revtex4-1}
\usepackage{graphicx}
\usepackage{amssymb}
\usepackage{amsmath}
\pdfoutput=1

\begin{document}

\def\e{\varepsilon}
\def\f{\varphi}
\def\p{\partial}
\def\ba{\mathbf{a}}
\def\bA{\mathbf{A}}
\def\bb{\mathbf{b}}
\def\bB{\mathbf{B}}
\def\bD{\mathbf{D}}
\def\be{\mathbf{e}}
\def\bE{\mathbf{E}}
\def\bH{\mathbf{H}}
\def\bj{\mathbf{j}}
\def\bk{\mathbf{k}}
\def\bK{\mathbf{K}}
\def\bM{\mathbf{M}}
\def\bm{\mathbf{m}}
\def\bn{\mathbf{n}}
\def\bq{\mathbf{q}}
\def\bp{\mathbf{p}}
\def\bP{\mathbf{P}}
\def\br{\mathbf{r}}
\def\bR{\mathbf{R}}
\def\bS{\mathbf{S}}
\def\bu{\mathbf{u}}
\def\bv{\mathbf{v}}
\def\bV{\mathbf{V}}
\def\bw{\mathbf{w}}
\def\bx{\mathbf{x}}
\def\by{\mathbf{y}}
\def\bz{\mathbf{z}}
\def\Bn{\boldsymbol{\nabla}}
\def\Bo{\boldsymbol{\omega}}
\def\Br{\boldsymbol{\rho}}
\def\Bs{\boldsymbol{\hat{\sigma}}}
\def\bh{{\beta\hbar}}
\def\mA{\mathcal{A}}
\def\mB{\mathcal{B}}
\def\mD{\mathcal{D}}
\def\mF{\mathcal{F}}
\def\mG{\mathcal{G}}
\def\mH{\mathcal{H}}
\def\mI{\mathcal{I}}
\def\mL{\mathcal{L}}
\def\mO{\mathcal{O}}
\def\mP{\mathcal{P}}
\def\mT{\mathcal{T}}
\def\mU{\mathcal{U}}
\def\mZ{\mathcal{Z}}
\def\fr{\mathfrak{r}}
\def\ft{\mathfrak{t}}
\newcommand{\rf}[1]{(\ref{#1})}
\newcommand{\al}[1]{\begin{aligned}#1\end{aligned}}
\newcommand{\ar}[2]{\begin{array}{#1}#2\end{array}}
\newcommand{\eq}[1]{\begin{equation}#1\end{equation}}
\newcommand{\bra}[1]{\langle{#1}|}
\newcommand{\ket}[1]{|{#1}\rangle}
\newcommand{\av}[1]{\langle{#1}\rangle}
\newcommand{\AV}[1]{\left\langle{#1}\right\rangle}
\newcommand{\braket}[2]{\langle{#1}|{#2}\rangle}
\newcommand{\ff}[4]{\parbox{#1mm}{\begin{center}\begin{fmfgraph*}(#2,#3)#4\end{fmfgraph*}\end{center}}}

\def\mr{m_{\perp}}
\def\ml{m_{\parallel}}
\def\hr{H_{\perp}}
\def\hl{H_{\parallel}}

\def\mb{(\mu+\alpha\nu)}
\def\nb{(\nu-\alpha\mu)}
\def\lb{(\lambda+\alpha\kappa)}
\def\kb{(\kappa-\alpha\lambda)}
\def\mn{\left|\bm\times\bz\right|}

\title{Nonlinear Dynamics in a Magnetic Josephson Junction}

\author{Silas Hoffman}
\affiliation{Department of Physics and Astronomy, University of California, Los Angeles, California 90095, USA}

\author{Ya.~M. Blanter}
\affiliation{Kavli Institute of Nanoscience, Delft University of Technology, 2628 CJ Delft, The Netherlands}

\author{Yaroslav Tserkovnyak}
\affiliation{Department of Physics and Astronomy, University of California, Los Angeles, California 90095, USA}

\begin{abstract}
We theoretically consider a Josephson junction formed by a ferromagnetic spacer with a strong spin-orbit interaction or a magnetic spin valve, i.e., a bilayer with one static and one free layer. Electron spin transport facilitates a nonlinear dynamical coupling between the magnetic moment and charge current, which consists of normal and superfluid components. By phenomenologically adding reactive and dissipative interactions (guided by structural and Onsager symmetries), we construct magnetic torques and charge pumping, whose microscopic origins are also discussed.  A stability analysis of our coupled nonlinear systems generates a rich phase diagram with fixed points, limit cycles, and quasiperiodic  states. Our findings reduce to the known  phase diagrams for current-biased nonmagnetic Josephson junctions, on the one hand, and spin-torque driven magnetic films, on the other, in the absence of coupling between the magnetic and superconducting order parameters.
\end{abstract}



\maketitle

Hybrid structures with ferromagnet (F)$\mid$normal-metal (N) interfaces have garnered much attention over the past few decades owing to their application in spintronic devices.  Injecting a spin current into such a system exerts a torque on the magnet \cite{slonczewskiJMMM96,*bergerPRB96}, which can induce precession and even reversal \cite{tsoiPRL98,*myersSCI99}, allowing for manipulation of the magnetic order parameter in nanoscale structures without an external magnetic field \cite{[{}][{, and references therein. }]ralphJMMM08}.  Because of the nonlinear nature of the ensuing magnetic dynamics, such devices offer observation of effects traditionally seen in nonlinear dynamical systems: Phase locking, hysteresis, bifurcations, and chaos are readily observed \cite{bertottiBOOK09}.

In consideration of a superconductor (S)$\mid$F$\mid$S heterostructure, one may expect the Josephson effect to be suppressed due to the rapid decay of a singlet pair inside the ferromagnet.  Recent experiments \cite{keizerNAT06,*khairePRL10,*robinsonSCI10}, however, observed superconducting transport through a strong ferromagnet between two conventional ($s$-wave) superconductors. With the expectation that the triplet component of the superconducting condensate can penetrate long lengths into a ferromagnet, the preservation of this signal suggests a spin singlet-to-triplet conversion at the interfaces \cite{[{}][{, and references therein. }]bergeretRMP05}. The unexpected persistence of a supercurrent through the magnet forecasts a new kind of spintronic device that manipulates the Josephson junction by the ferromagnet and, conversely, ferromagnetic layer by the superconducting condensate \cite{braudePRL08,*konschellePRL09,waintalPRB02,*linderPRB11,petkovicPRB09,*caiPRB10,*barnesSST11}.

Previous analyses \cite{buzdinPRL08,braudePRL08,linderPRB11,petkovicPRB09} have considered \textit{equilibrium} interactions between magnetic and superconducting order parameters, which naturally induce a reactive coupling.  In contrast, in our description, we introduce \textsl{nonequilibrium} interactions consistent with the symmetries of the structure and obeying Onsager reciprocity \cite{landauBOOKv5}.  This treatment allows the addition of	 both dissipative and reactive couplings between the magnet and superconductor that may in practice be crucial in the understanding of ferromagnetic Josephson junctions, analogous to the importance of Slonczewski \cite{slonczewskiJMMM96,tsoiPRL98,ralphJMMM08} and spin-pumping \cite{[{}][{, and references therein. }]tserkovRMP05} terms in the theory of spin-transfer torques.  Such effects cannot be fully captured by quasiequilibrium free-energy considerations. We expect the dissipation to be governed by the quasiparticle excitations in the superconductors in concert with the microscopic processes in the ferromagnet (such as magnon-phonon and magnon-magnon interactions in insulators) that are responsible for their Gilbert damping (which, in turn, is known to persist down to very low temperatures).

\begin{figure}[pt]
\includegraphics[width=\linewidth,clip=]{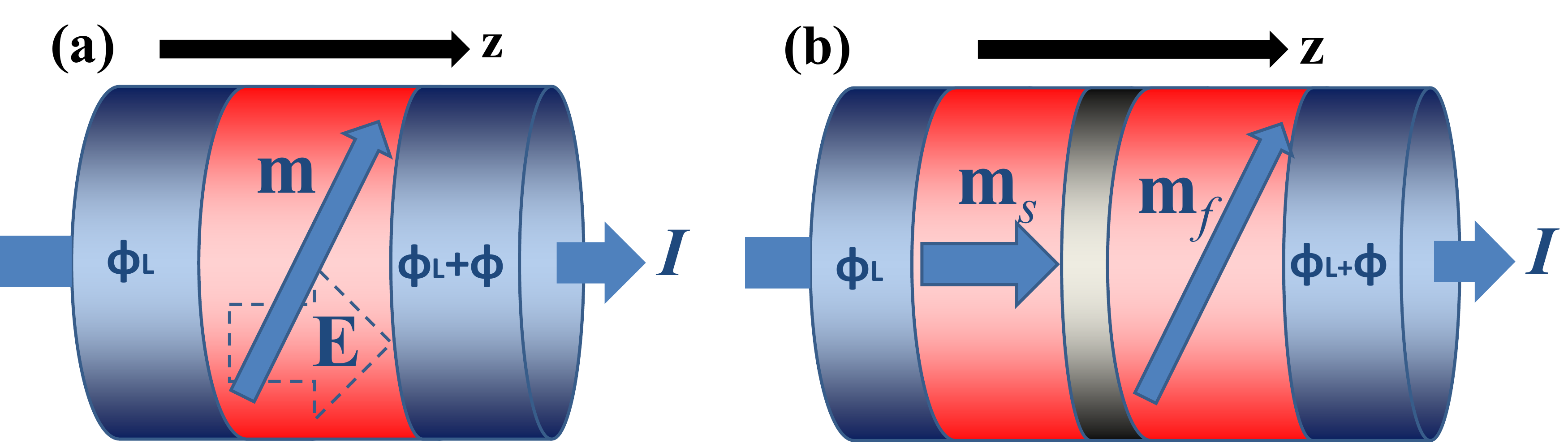}
\caption{Schematics of our magnetic Josephson junctions. The directions of junction layering, applied current $I$, internal Rashba field $\mathbf{E}$ (a), and direction of the static ferromagnetic layer $\textbf{m}_s$ (b) all lie along the $\bz$ axis. $\phi$ is the phase difference between the superconducting leads.}
\label{device}
\end{figure}

In order to provide specific examples, we consider (a) an S$\mid$F$\mid$S heterostructure with a Rashba spin-orbit interaction (SOI) in a thin ferromagnetic interlayer (neglecting the vector potential and associated phase shift caused by its magnetic moment \cite{caiPRB10}) and (b) S$\mid$F$\mid$N$\mid$F$\mid$S heterostructure wherein one ferromagnetic layer is pinned, $\bm_s$, and the other free, $\bm_f$. See Fig.~\ref{device}. The corresponding spin-dependent Hamiltonians mix the singlet and triplet superconducting components \cite{bergeretRMP05}, allowing the superfluid to penetrate into the magnet and exert spin torque and carry spin pumping (since a triplet Cooper pair is a spin-1 object) that are analogous to those associated with normal quasiparticles (spin-$1/2$ objects). In particular, as a simple model to demonstrate proof of concept, we take the device geometry to be rotationally symmetric along the axis associated with the Rashba interaction, as sketched in Fig.~\ref{device}(a), and along the direction of the fixed magnetic layer, as sketched in Fig.~\ref{device}(b). By analyzing the stability and dynamics of our model, we outline a phase diagram of the coupled system as a function of applied magnetic field and current bias.

The phenomenological equation of motion of an isolated ferromagnet sufficiently well below the Curie temperature is given by the Landau-Lifshitz-Gilbert (LLG) equation \cite{landauBOOKv9,*gilbertIEEEM04}
\eq{\dot{\bm}=-\gamma\bm\times\bH+\alpha\bm\times\dot{\bm}\,,}
where $\gamma$ is the gyromagnetic ratio and $\alpha$ is the dimensionless Gilbert damping.  We use a normalized form of this equation, in which the (unit) magnetic direction vector $\bm=\textbf{M}/M_s$, $M_s=\left|\textbf{M}\right|$ (saturation magnetization), is dimensionless.  $\bH=-\mathcal{V}^{-1}\partial F/\partial \textbf{M}$ is the effective magnetic field and $F$, $\textbf{M}$, and $\mathcal{V}$ are the free energy, magnetization vector, and volume, respectively. In the spin-valve model, Fig.~\ref{device}(b), $\bm$ will denote the free layer $\bm_f$.

We consider the resistively-shunted junction (RSJ) model for the Josephson junction, wherein the device is composed of conventional superconductors with some Ohmic conductance $\sigma$ in the junction \cite{likharevBOOK86}. Additionally, we take the capacitance to be zero, which precludes $RC$-type delays in the coupled dynamics. The corresponding Josephson relations (for a static magnetization) are
\eq{\dot{Q}=I_c\sin\phi+\sigma V\,,\,\,\,\dot{\phi}=\frac{2e}{\hbar}V\,,\label{josephRel}}
where $V$ is the voltage drop across the junction. $\phi$ is the phase difference between the superconducting reservoirs and $Q$ is the charge transported by the junction. The supercurrent is proportional to the critical current, $I_c=(2e/\hbar)E_J$, where $E_J$ parametrizes the Josephson energy $-E_J\cos\phi$. We note that Eq.~\eqref{josephRel} is dictated by gauge symmetry and, in anticipation of the arguments to follow, is a manifestation of Onsager reciprocity in the dynamics of $Q$ and $\phi$.

Under time reversal, $\dot{\bm}\rightarrow\dot{\bm}$ and $\alpha\bm\times\dot{\bm}\rightarrow-\alpha\bm\times\dot{\bm}$.  The term proportional to $\alpha$ in the LLG equation thus reflects irreversible processes.  We characterize such terms as dissipative.  $\sigma$, likewise, parametrizes Ohmic dissipation of normal fluid.  All other terms thus considered so far are reactive. Couplings between the free ferromagnet and superconductor at the level of the free energy, induced by the static magnetic layer or SOI, are restricted by the symmetries of our structure.  Our device geometries shown in Fig.~\ref{device} are invariant under rotations about the $\bz$ axis as well as certain combinations of discrete symmetries. Because both the exchange interaction between the magnetic layers of our spin-valve device and the Josephson energy are individually preserved under the symmetries of the combined system, the product of these interactions must also be permitted \cite{linderPRB11}. However, the interlayer F$\mid$N$\mid$F spin-valve exchange is usually very small (except for the thinnest N spacers) \cite{ralphJMMM08}, and will be disregarded in our study.  One may, furthermore, show that any (time-reversal symmetric) quadratic cross term involving $\bm$, $Q$, and $\phi$ does not respect the symmetries of our device geometry (keeping in mind that $\phi\to-\phi$ under time reversal and $\mathbf{m}$ is a pseudovector under improper rotations). In particular, an interaction of the type $\cos(\phi+\Gamma m_z)$ \cite{buzdinPRL08} is forbidden in our geometry. Thus neglecting interactions of $\bm$, $Q$, and $\phi$ beyond quadratic order, the free energy remains uncoupled:
\eq{F[\bm,Q,\phi]=F[\bm]+F[Q]+F[\phi]\,,}
where $F[\bm]=\mathcal{V}KM_z^2/2-\mathcal{V}\textbf{M}\cdot\textbf{H}_a$, $F[Q]=-QV$, and $F[\phi]=-E_J\cos\phi$.  The sign of the anisotropy constant, $K$, defines an easy plane or easy axis and is determined by the geometry of the device and crystalline anisotropies. $\textbf{H}_a$ is an applied external magnetic field.

The LLG equation of motion of the magnet is now complemented with interactions that are quasistationary (i.e., first order in frequency), up to quadratic order in the components of $\textbf{m}$, preserving the magnitude of $\bm$, and consistent with the structural symmetry of the device:
\begin{align}
\dot{\bm}=&-\gamma\bm\times\bH+\alpha\bm\times\dot{\bm}\nonumber\\
&+(\mu\dot{Q}+\lambda\dot{\phi})\bm\times\bz\times\bm+(\nu\dot{Q}+\kappa\dot{\phi})\bm\times\bz\,.
\label{LLGg}
\end{align}
Hereafter, we are focusing on the spin-valve case, Fig.~\ref{device}(b), where the phenomenological coupling coefficients $\mu$, $\lambda$, $\nu$, and $\kappa$ may be taken to be angle-independent constants and $\mathbf{z}=\bm_s$. [For the SOI device, Fig.~\ref{device}(a), structural symmetries dictate these coefficients to be odd functions in $m_z$.] Constants $\mu$ and $\nu$ characterize the strength of the coupling between the magnet and the total current $\dot{Q}$. Similarly, the strength of the coupling between the magnet and the dynamics of the superfluid condensate $\dot{\phi}$ is characterized by $\lambda$ and $\kappa$. To the reader familiar with spin valves \cite{ralphJMMM08}, Eq.~(\ref{LLGg}) is reminiscent of the Landau-Lifshitz-Gilbert equation with the so-called Slonczewski and field-like torques, respectively, added on the second line of the right-hand side. In this case, sketched in Fig.~\ref{device}(b), current is spin polarized by passing through the fixed magnetic layer. The resulting spin-polarized current impinging on a free ferromagnet induces torque due to conservation of angular momentum. In the case of a single magnetic layer with SOI, Fig.~\ref{device}(a), a spin torque is generated via SOI inside this layer itself \cite{manchonPRB08}. Because the leads in our system are superconducting, we additionally generate a torque as a result of the dynamics of the superfluid condensate.  Loosely speaking, the torque induced by both currents, normal current and supercurrent, through the junction produce two channels for driving magnetization dynamics (and thus two sets of terms, as compared to the usual normal-metal spin torques).  Appropriately, above the critical temperature of the superconductor, we expect to recover the normal-metal limit, in which torque is generated by the ordinary Ohmic current alone.

The reaction of the current and superconducting phase dynamics to the magnet are not captured by the Josephson relations, Eq.~(\ref{josephRel}), which would not be consistent with Eq.~(\ref{LLGg}). One must extend Eq.~\eqref{josephRel} to include the pumping terms satisfying Onsager reciprocity, in order to obtain equations of motion for our coupled system that obey microscopic time-reversal symmetry \cite{landauBOOKv5}. Because the magnet flips under time reversal (upon invoking Onsager symmetry), one must additionally use the symmetries of the structure to relate the time-reversed state to the original.  After straightforward manipulations, that are  analogous to Ref.~\cite{halsEPL10} for normal junctions, we construct the following equations in lieu of Eq.~\eqref{josephRel}:
\begin{align}
\dot{Q}&=\frac{2e}{\hbar}\left[E_J\sin\phi-\mathcal{S}(\lambda\dot{\bm}\cdot\bm\times\bz+\kappa\dot{\bm}\cdot\bz)\right]+\frac{\hbar\sigma}{2e}\dot{\phi}\,,\nonumber\\
\dot{\phi}&=\frac{2e}{\hbar}\left[V-\mathcal{S}(\mu\dot{\bm}\cdot\bm\times\bz+\nu\dot{\bm}\cdot\bz)\right]-\rho\dot{Q}\,,
\end{align}
where $\mathcal{S}=\mathcal{V}M_s/\gamma$ is the total spin angular momentum of the ferromagnetic layer. These equations of motion now include both normal and superfluid pumping, which are Onsager reciprocal to the driving effects introduced in the generalized LLG equation, Eq.~(\ref{LLGg}).  Our theory includes two types of pumping as a result of the non-Ohmic relationship between current and voltage. The term with coefficient $\rho$ causes current to drag phase across the device; $\rho$ is a measure of the viscosity between the current and superfluid condensate.  Although this term is not needed for consistency with Onsager reciprocity, we will see that it would generally have to be included in order to satisfy the second law of thermodynamics. We could also immediately notice that the coefficients $\rho$, $\nu$, and $\mu$ should vanish in the limit of large superconducting reservoirs, recovering the ordinary ac Josephson effect (as expected based on the gauge invariance). Keeping these terms, on the other hand, would capture finite-size (mesoscopic) properties of the superconducting layers, which are of secondary interest to our ends.

We may write the equations of motion in a dimensionless form by measuring time, magnetic field, charge, voltage, and conductance in units of $\mathcal{S}/E_J$, $E_J/\gamma\mathcal{S}$, $2e\mathcal{S}/\hbar$, $E_J\hbar/2e\mathcal{S}$, and $\mathcal{S}(2e/\hbar)^2$, respectively:
\begin{align}
\dot{\bm}=&-\bm\times\textbf{H}+\alpha\bm\times\dot{\bm}+\dot{\phi}(\lambda\bm\times\bz\times\bm+\kappa\bm\times\bz)\nonumber\\
&+\dot{Q}(\mu\bm\times\bz\times\bm+\nu\bm\times\bz)\,,\nonumber\\
\dot{Q}=&\sin\phi-\lambda\dot{\bm}\cdot\bm\times\bz-\kappa\dot{\bm}\cdot\bz+\sigma\dot{\phi}\,,\nonumber\\
\dot{\phi}=&V-\mu\dot{\bm}\cdot\bm\times\bz-\nu\dot{\bm}\cdot\bz-\rho\dot{Q}\,.
\label{gen}
\end{align}
Additionally, allow us to absorb a factor of $\mathcal{V}M_s^2/E_J$ into the anisotropy constant, such that the free energy for the magnet reads $F[\bm,Q,\phi]=E_J(Km_z^2/2-\bm\cdot\textbf{H}_a-QV-\cos\phi)$.  Under time reversal, the terms with coefficients $\nu$ and $\lambda$ reverse sign in the LLG equation.  Because $\dot{\bm}$ does not change sign, these are dissipative.  Likewise, the terms with coefficients $\mu$ and $\kappa$ do not reverse sign and are thus nondissipative.  $\sigma$ is a dissipative coefficient, therefore $\rho$ is as well.

Let us try to understand the microscopic origin of the dissipative terms in our theory.  Consider momentarily only the RSJ subsystem: when enough energy is supplied (either thermally or by a bias), quasiparticles are able to overcome the superconducting gap and transport through the junction. Normal scattering of quasiparticles across the junction causes Ohmic resistance.  Likewise, consider an isolated precessing ferromagnet. This is microscopically described by a coherent magnon state that can decay into phonons and incoherent magnons, processes which macroscopically give Gilbert damping.  In the case of a metallic ferromagnet, the additional decay channel into the electron-hole continuum enhances further its Gilbert damping. Upon coupling these subsystems, energy is shared by the entire structure. Likewise, dissipation by microscopic mechanisms underlying Ohmic conductance and Gilbert damping can give rise to a dissipative (viscous) energy transfer between ferromagnetic and superconducting layers, as parametrized by new dissipative coefficients $\nu$ and $\lambda$. Phenomenologically, therefore, we may expect $\sigma$, $\alpha$, and $\rho$ to bound $\nu$ and $\lambda$, which is indeed verified below.

In the RSJ model, Eq.~\eqref{josephRel}, if $\phi$ is static, we are in a superconducting (S) state because only dissipationless current is passing through the junction.  Likewise if $\phi$ is not constant, the circuit must have a finite voltage drop. This is called a resistive (R) state. Notice that in our generalized model, Eqs.~\eqref{gen}, a choice of dynamics that leave $\phi$ static can still generate dissipative current due to magnetic pumping. We will, nonetheless, keep refering to the static and dynamic states of $\phi$ as the superconducting (S) and resistive (R) states, respectively, even though this terminology is, in general, abusive, in the presence of the new spin-torque/pumping terms in Eqs.~\eqref{gen}.

We distinguish between two regimes governed by the superconducting coherence length $\xi$.  When $\xi$ is smaller than the width of superconducting terminals, the bulk properties of the superconductors will be largely detached from physics at the interfaces.  Thus for large superconducting reservoirs, a change in phase difference cannot be induced by transport through the junction.  We expect the corresponding coefficients $\mu$, $\nu$, and $\rho$ to scale inversely with the volume of the smaller of the superconducting layers then; these are representative of mesoscopic effects, as has already been inferred above.  Because charge is a conserved hydrodynamic quantity, on the other hand, there is a length at which the dynamics at the interface decouple from the current in the bulk.  In particular, $\kappa$, $\lambda$, and $\sigma$ should not depend on the size of a superconducting reservoir; these coefficients parametrize the properties of the Josephson junction itself and are thus of central interest to us. Dynamic properties of mesoscopic junctions with nonzero $\nu$ and $\mu$ terms in Eqs.~\eqref{gen} are discussed in the Supplementary Text, where, in particular, we point out a reentrant behavior for the resistive (R) state as a function of the applied current $I$.

For our analysis, it is convenient to consider dc current biasing (setting $\rho$, $\nu$, $\mu$ to zero), $\dot{Q}=I$:
\begin{align}
\dot{\bm}&=-\bm\times\textbf{H}+\alpha\bm\times\dot{\bm}+\dot{\phi}(\lambda\bm\times\bz\times\bm+\kappa\bm\times\bz)\,,\nonumber\\
\sigma\dot{\phi}&=I-\sin\phi+\lambda\dot{\bm}\cdot\bm\times\bz+\kappa\dot{\bm}\cdot\bz\,.
\label{fullCurrent}
\end{align}
In what follows, we treat a system where the applied magnetic field is along the axis of symmetry, $\textbf{H}_a=H_a\hat\bz$, and $K$ is positive (which is generically the case for films with magnetostatic energy dominating over crystalline anisotropy). Thermodynamic self consistency of our theory requires for the dissipation power  $P=(E_J^2/\mathcal{S})(\alpha\dot\bm^2-2\lambda\dot{\bm}\cdot\bm\times\bz\dot\phi+\sigma\dot{\phi}^2)\geq0$. This bounds our phenomenological constant $\lambda$ as $\lambda^2\leq\alpha\sigma$ (while, clearly, $\alpha\geq0$ and $\sigma\geq0$). To proceed with the nonlinear stability analysis, it is important to notice that, according to Eqs.~\eqref{fullCurrent}, the dynamics of $m_z$ and $\phi$ decouple from the transverse magnetization $(m_x,m_y)$, which can, in turn, be expressed in terms of $(m_z,\phi)$ [see Eq.~\eqref{transSol} in the Supplementary Text].

\begin{figure}[pt]
\includegraphics[width=\linewidth,clip=]{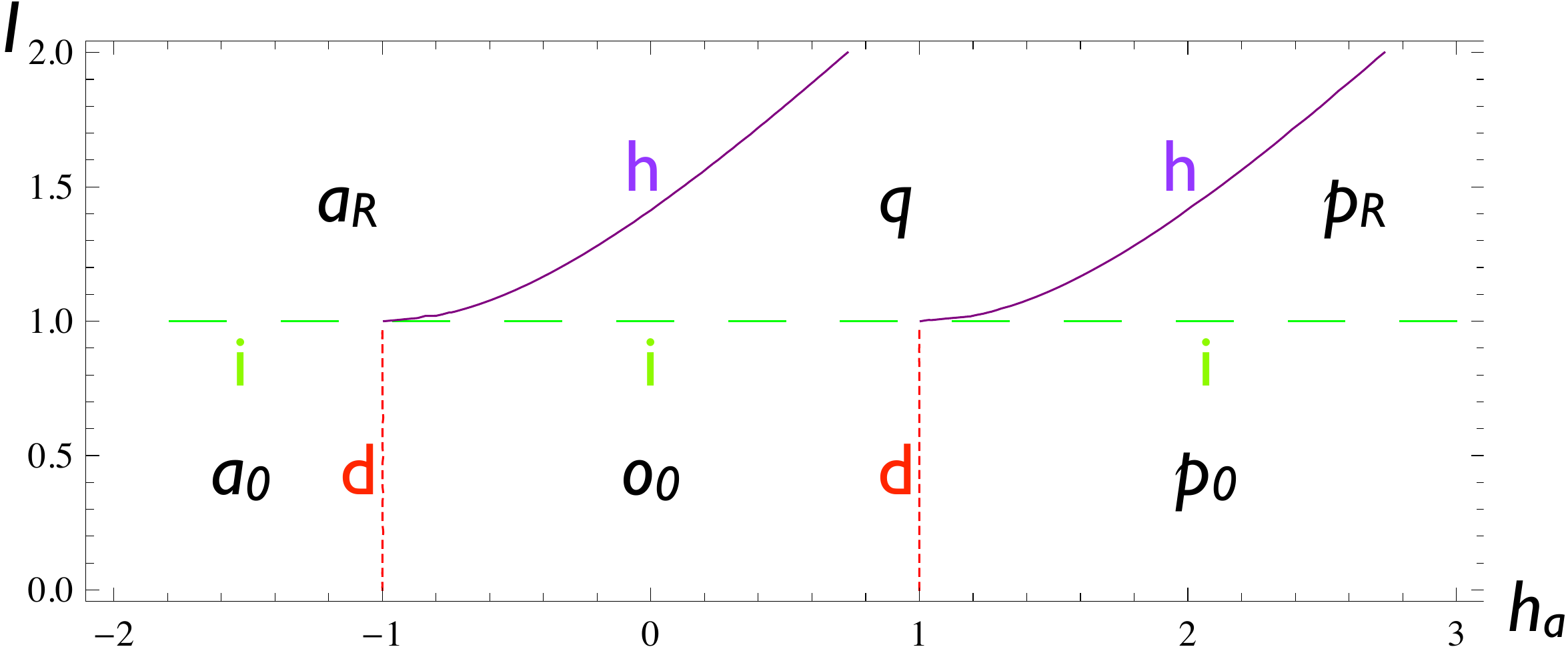}
\caption{Stability diagram as a function of the current, $I$, and applied magnetic field $h_a$.  $\lambda=-0.1$, $\mu,\nu,\kappa=0$, $K=1$, $\alpha=1$, and $\sigma=0.1$. h labels the Hopf bifurcation (solid lines), i labels the infinite-period bifurcation (long-dashed lines), and d labels the saddle-node bifurcation (short-dashed lines).}
\label{fig2}
\end{figure}

When the current is below the critical current, $I\leq 1$, one can show that there are three stable fixed points: $p_0$, $a_0$, and $o_0$ which correspond to a Josephson 0-junction (defining a junction with $\left|\phi\right|<\pi/2$ to be in the ``0 phase" and $\pi/2<\left|\phi\right|<\pi$ in the ``$\pi$ phase") and magnetic direction parallel, antiparallel, and away from the $z$ axis, respectively. In all these states $\phi$ is fixed by the applied current such that $\sin\phi=I$.  As indicated by our stability diagram, Fig.~\ref{fig2}, the state of our device is determined by the applied magnetic field.  When $|h_a|\leq 1$, where $h_a\equiv H_a/K$, $m_z=h_a$ and $m$ is fixed by initial conditions.  By applying a sufficiently large external magnetic field, $|h_a|\geq 1$, $o_0$ is annihilated under a saddle-node bifurcation \cite{guckenheimerBK83}, and the sole stable state is $p_0$ or $a_0$ for positive or negative applied field, respectively, pinning the magnet along the $z$ axis.  A full linear stability analysis is discussed in the Supplementary Text where we note, specifically, that the dissipation power bound precludes the existence of a $\pi$-junction.

If the current exceeds its critical value, $I>1$, the superconducting phase and $z$ component of the magnet become dynamic.  This disappearance of all the fixed points is an infinite-period bifurcation \cite{guckenheimerBK83}.  Because no fixed points exist, the Poincar{\'e}-Bendixson theorem implies any closed orbit on the cylinder, parameterized by $m_z$ and $\phi$, is periodic and must go around the circumference of this cylinder.  Supposing the  frequency of this periodic motion is $\Omega_J$, $m_z$ may be written as a constant plus terms periodic in $\Omega_J$.  Likewise, we may express $\phi=n\Omega_J t$ (with nonzero $n\in\mathbb{Z}$) plus terms periodic in $\Omega_J$. Therefore, the characteristic frequency of the system is given by the time average of $\dot{\phi}$. Upon solving the equation of motion for transverse component of the magnet [Eq.~\eqref{transSol} in the Supplementary Text], we find it undergoes rotations at frequency $n(\lambda/\alpha)\Omega_J$ that are superimposed with $\Omega_J$ oscillations.  Therefore the magnet in general undergoes quasiperiodic motion, a state we label $q$.

To determine the full expression for $\Omega_J$ when $I>1$ would require solving the system of differential equations \eqref{fullCurrent}. For simplicity, consider the limit of small $\lambda$ and $\kappa$, so that we can neglect quadratic terms in $\lambda$ and $\kappa$ in Eqs.~(\ref{fullCurrent}). In this case, the characteristic frequency of the Josephson junction is given by the usual RSJ frequency $\omega_J=\sqrt{I^2-1}/\sigma$ \cite{likharevBOOK86}.  In region $q$ of our stability diagram, Fig.~\ref{fig2}, $m_z$ oscillates with frequency $\omega_J$ around the average value $\mbox{sign}(I)(\lambda/\alpha-\kappa)\omega_J+h_a/K$.  Near the point $\left|\mbox{sign}(I)(\lambda/\alpha-\kappa) \omega_J+h_a/K\right|=1$, a Hopf bifurcation \cite{guckenheimerBK83} (labeled h) is induced wherein the quasiperiodic orbit disappears and the magnet is parallel or antiparallel to the $z$ axis, labeled $p_R$ and $a_R$ respectively, and the phase is dynamic. We anticipate the higher-order coupling in $\lambda$ and $\kappa$ to modify the frequency dependence on current. Furthermore, we expect  that, near the line defining the Hopf bifurcation, there exists a phase of bimodal stability wherein the magnet can orient along the $z$ axis or precess quasiperiodically, subject to the initial conditions. This is a natural consequence of the reciprocity of current-driven magnetic dynamics and pumping and persists even in the absence of any superconductivity (i.e., $E_J=0$). Details of these rich coupled nonlinear dynamics are, however, beyond the scope of the present paper. 

In summary, we have introduced a model of S$\mid$F$\mid$S and S$\mid$F$\mid$N$\mid$F$\mid$S heterostructures coupling the dynamics of the magnets to that of the superconductor via a Rashba SOI in single-layer junctions and via magnetic misalignment in spin-valve junctions. We expect such structures to be highly adaptable to uses in spintronics due to the versatility with which one can in principle influence both the magnet and superconductor. Chaos in ferrites and magnetic thin films is often attributed to spatially nonuniform magnetizations \cite{bertottiBOOK09}. Perhaps a simpler route towards chaos in our model is by applying a magnetic field perpendicular to the axis of cylindrical symmetry. As a result, the dynamic equations become three dimensional and thus no longer restricted by the Poincar{\'e}-Bendixson theorem to periodic orbits or fixed points.

\clearpage

\appendix

\begin{widetext}
\section{Supplementary Text}

\textit{Decoupled Junction.}|In the special case where mesoscopic effects dominate ($\mu,\nu,\rho\neq0$ but $\lambda,\kappa$=0) in Eqs.~(6), the current-biased magnetic and superconducting dynamics decouple.  We take this opportunity to recall the properties of magnetic spin valves and the RSJ model of superconductors, to which the decoupled equations map. Ignoring $\lambda$ and $\kappa$, Eqs.~(6) become simply
\begin{equation}
\dot{\bm}=-\bm\times\textbf{H}+\alpha\bm\times\dot{\bm}+I(\mu\bm\times\bz\times\bm+\nu\bm\times\bz)\,,\,\,\,\sigma\dot{\phi}=I-\sin\phi\,.
\end{equation}
The equation of motion for the magnet is thus the LLG equation for a spin valve, including Slonczewski ($\mu$) and field-like ($\nu$) torques, in the case that a fixed magnetic layer points along the $z$ axis.  The superconductor is described by the RSJ model with zero capacitance. Appropriately, we find that the dissipation power depends only on the dissipative constants: $P=(E_J^2/\mathcal{S})(\alpha\dot\bm^2+2\nu\dot{\bm}\cdot\bz\dot Q+\sigma\dot{\phi}^2+\rho\dot Q^2)\geq0$. Because $\alpha\geq0$, $\sigma\geq0$, and $\rho\geq0$, our phenomenological constant is bounded, $\nu^2\leq\alpha\rho$, as anticipated. There are three possible stable states of the current-biased magnet in the presence of a static field in the $z$ direction: pinned parallel to the $z$ axis, antiparallel to the $z$ axis, or precessing around the $z$ axis, labeled $p$, $a$, and $o$, respectively. A pinned state is stable when $|(\mu/\alpha-\nu)I/K+h_a|\geq 1$.  If $|(\mu/\alpha-\nu)I/K+h_a|<1$, the magnet precesses at frequency $\omega_M=\mu I/\alpha$.  The corresponding stability diagram with Hopf bifurcation lines is shown in Fig.~\ref{fig3}. In the dimensionless form of the RSJ description, when $-1\leq I\leq 1$, the junction is in the S state and the phase is fixed at $\phi=\sin^{-1}I$.  When the current is raised beyond the critical current, $I>1$, the Josephson junction is in the R state and $\phi$ oscillates with  frequency $\omega_J=\sqrt{I^2-1}/\sigma$. For the RSJ model, a $\pi$ junction is trivially impossible: $\left|\phi\right|$ cannot access values between $\pi/2$ and $\pi$.  The inset of Fig.~\ref{fig3} displays the well-known phase diagram of the RSJ junction.\\

\begin{figure}[ph]
\includegraphics[width=0.49\linewidth,clip=]{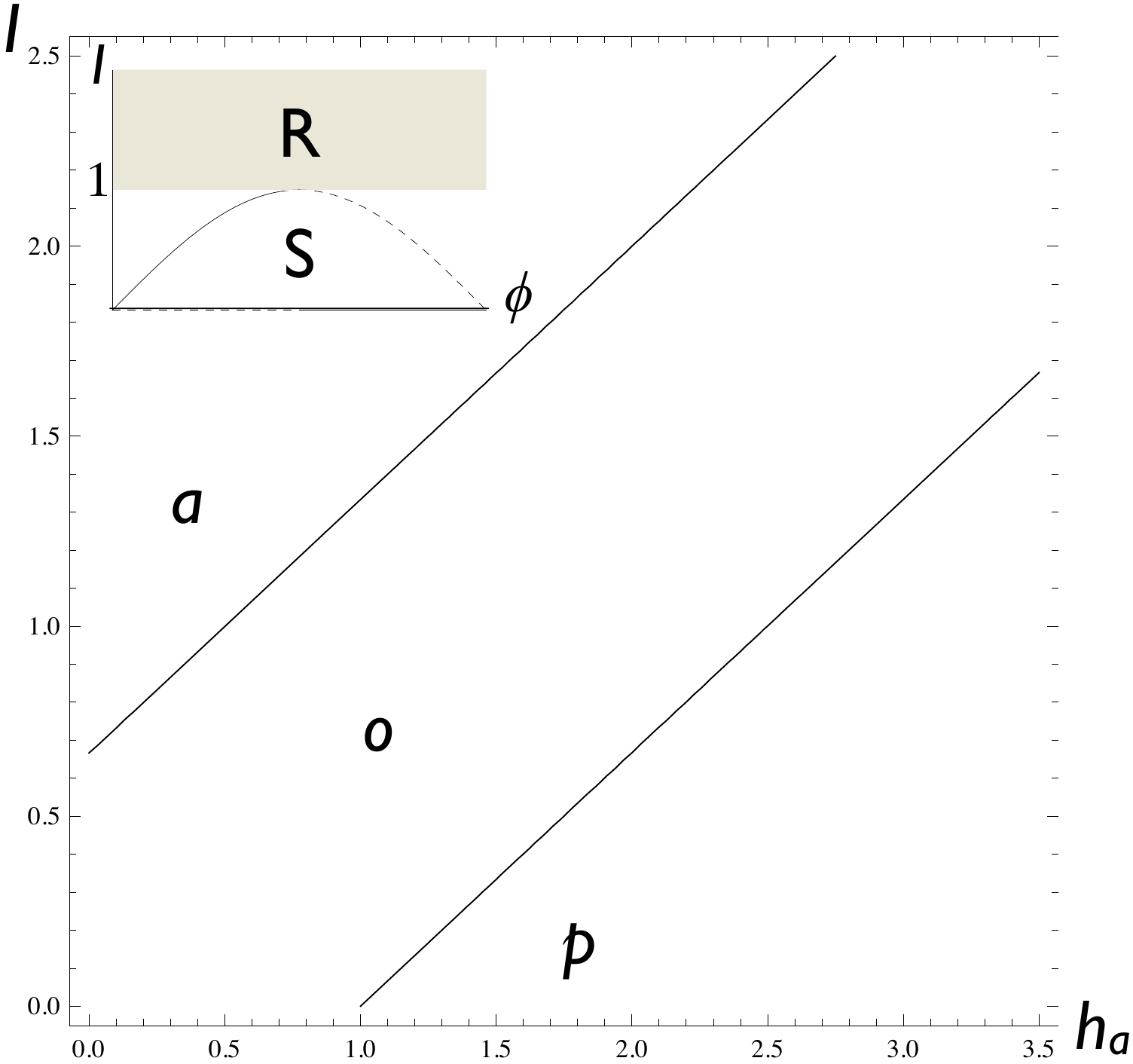}
\caption{Stability diagram as a function of the current and applied magnetic field of the decoupled magnet.  $\mu=-1.5$, $\nu,\lambda,\kappa=0$, and $K=1$. $p$ and $a$ label the parallel and antiparallel states of the magnet, respectively.  Inset: decoupled Josephson junction. The S state (unshaded) and R state (shaded) are separated by the line $I=1$. Solid line is the value of $\phi$ for a 0 junction and dashed for the unstable $\pi$ junction.}
\label{fig3}
\end{figure}

\textit{Stability Analysis}.|We enumerate all of the fixed points of the junction discussed in the Main Text and calculate their associated stability. The equations of motion for $m_z$ and $\phi$ decouple from the transverse dynamics, according to Eq.~(7):
\begin{equation}
\dot{m}_z=(1-m_z^2)\left[\bar\alpha(H_a- K m_z)+\bar\lambda \dot{\phi}\right]\,,\,\,\,\dot{\phi}=\frac{I-\sin\phi-\bar\lambda (H_a-K m_z)(1-m_z^2)}{\sigma-(\lambda\bar\kappa+\kappa\bar\lambda)(1-m_z^2)}\,,
\label{mzfDec}
\end{equation}
where $\bar\lambda\equiv(\lambda-\alpha\kappa)/(1+\alpha^2)$, $\bar\kappa\equiv(\kappa+\alpha\lambda)/(1+\alpha^2)$, and $\bar\alpha\equiv\alpha/(1+\alpha^2)$. The solution for the transverse components of the magnet, $m=m_x+im_y$, as a function of $m_z$ and $\phi$ are given, in turn, by
\eq{m=\sqrt{1-m_z^2}\exp\left[-\frac{i}{\alpha}\left(\lambda \phi+\frac{1}{2}\ln\frac{1-m_z}{1+m_z}\right)+i\varphi\right]
\label{transSol}}
with $\varphi$ determined by initial conditions.  Consequently, the fixed points of the equations of motion for $m_z$ and $\phi$,  Eq.~(\ref{mzfDec}), immediately determine the state of the full system. When the current is below the critical current, $I\leq 1$, there are at most six fixed points for the $(m_z$,$\phi$) dynamics [Eqs.~\eqref{mzfDec}]:
\eq{
(\bar m_z,\bar\phi) = \left\{(\pm 1,\sin^{-1}I)\,,\,\,\,(\pm 1,\pi-\sin^{-1}I)\,,\,\,\,(h_a,\sin^{-1}I)\,,\,\,\,(h_a,\pi-\sin^{-1}I)\right\}
\label{fixed}
\,.}
We henceforth label the first four fixed points as $p_0$, $a_0$, $p_\pi$ and $a_\pi$.  At these points, the magnet points parallel ($p$) or antiparallel ($a$) to the $z$ axis, and the Josephson junction is a 0 or $\pi$ junction.  We refer to the last two fixed points in Eq.~\eqref{fixed} (which are only possible when $|h_a|<1$) as $o_0$ and $o_\pi$, with $m_z$ determined by the ratio of the applied field to the magnetic anisotropy. When $|h_a|\geq 1$, $o_0$ and $o_\pi$ are annihilated under a saddle-node bifurcation, and only the fixed points pinned along the $z$-axis, $p_0$, $a_0$, $p_\pi$ and $a_\pi$, remain.  At these fixed points, the superconducting phase and magnetic order decouple.  The stability analysis for $\phi$ then reduces to the RSJ model, resulting, therefore, in a 0 junction. Thus, the only stable points, when $|h_a|\geq 1$ and $I\leq 1$, are $p_0$ and $a_0$ subject to the direction of the applied field.  For intermediate values of the applied magnetic field, $\left|h_a\right|<1$, $p_0$, $a_0$, $p_\pi$ and $a_\pi$ are all unstable fixed points and the system is at $o_0$ or $o_\pi$.  We can see that if $\sigma<(\lambda\bar\kappa+\kappa\bar\lambda)(1-m_z^2)$ [such that the denominator in Eq.~\eqref{mzfDec} for $\dot{\phi}$ can become negative], the magnet is capable of sustaining the system in a $o_\pi$ state. Achieving a $\pi$ junction for an optimal choice of $\kappa$ requires $\lambda^2>\alpha\sigma/(1-m_z^2)$, in direct contradiction with the aforementioned dissipation power bound. Therefore, in this model, a $\pi$ junction is forbidden and, specifically when $I\leq 1$, the stable states of our system are $p_0$, $a_0$, or $o_0$, according to the value of $h_a$.\\

\textit{General Junction}.|In the following, we analyze the properties of the general junction wherein we do not restrict any phenomenological parameters in Eqs.~(6) to be zero. As previously, the transverse magnetization $m=m_x+im_y$ decouples from the $(m_z,\phi)$ dynamics:
\begin{equation}
\dot m_z =(1-m_z^2)\left[\bar\alpha(H_a-K m_z)+\bar\lambda\dot\phi+\bar\mu I\right]\,,\,\,\,\dot{\phi}=\frac{I-\sin\phi-\left[\bar\lambda(H_a-K m_z)-I(\bar\mu\kappa+\bar\nu\lambda)\right](1-m_z^2)}{\sigma-(\lambda\bar\kappa+\kappa\bar\lambda)(1-m_z^2)}\,,
\end{equation}
where $\bar\mu\equiv(\mu-\alpha\nu)/(1+\alpha^2)$, $\bar\nu\equiv(\nu+\alpha\mu)/(1+\alpha^2)$ and $\bar\lambda\equiv(\lambda-\alpha\kappa)/(1+\alpha^2)$, $\bar\kappa\equiv(\kappa+\alpha\lambda)/(1+\alpha^2)$, $\bar\alpha\equiv\alpha/(1+\alpha^2)$, as before. One can show that the general solution for transverse components is (up to an arbitrary phase shift $\varphi$)
\eq{m=\sqrt{1-m_z^2}\exp\left[-\frac{i}{\alpha}\left(\mu I t + \lambda \phi+\frac{1}{2}\ln\frac{1-m_z}{1+m_z}\right)\right]\,.}
The fixed points in the $(m_z,\phi)$ plane are
\eq{
(\bar m_z,\bar\phi) = \left\{(\pm 1,\sin^{-1}I)\,,\,\,\,(\pm 1,\pi-\sin^{-1}I)\,,\,\,\,\left((\mu/\alpha-\nu)I/K+h_a,\sin^{-1}I'\right)\,,\,\,\,\left((\mu/\alpha-\nu)I/K+h_a,\pi-\sin^{-1}I'\right)\right\}\,,
}
where we have introduced
\eq{I'\equiv I\left[1+(\mu/\alpha)\lambda\left(1-\bar{m}_z^2\right)\right]}
with $\bar{m}_z=(\mu/\alpha-\nu)I/K+h_a$ that itself depends on the current bias $I$. At the first four fixed points, the magnet is pinned parallel or antiparallel to the $z$ axis and can be either a 0 or $\pi$ junction.  Hence, maintaining consistent language between the coupled and general junctions, we label these fixed points $p_0$, $a_0$, $p_\pi$, and $a_\pi$. The final two fixed points [which are possible when $|(\mu/\alpha-\nu)I/K+h_a|<1$] are labeled by $o_0$ and $o_\pi$. These $o_0$ and $o_\pi$ points are stationary in the $(m_z,\phi)$ plane but the transverse components of the magnet follow a circular orbit of radius $\sqrt{1-\bar m_z^2}$ at frequency $\omega_M=\mu I/\alpha$.

\begin{figure}[pth]
\includegraphics[width=0.49\linewidth,clip=]{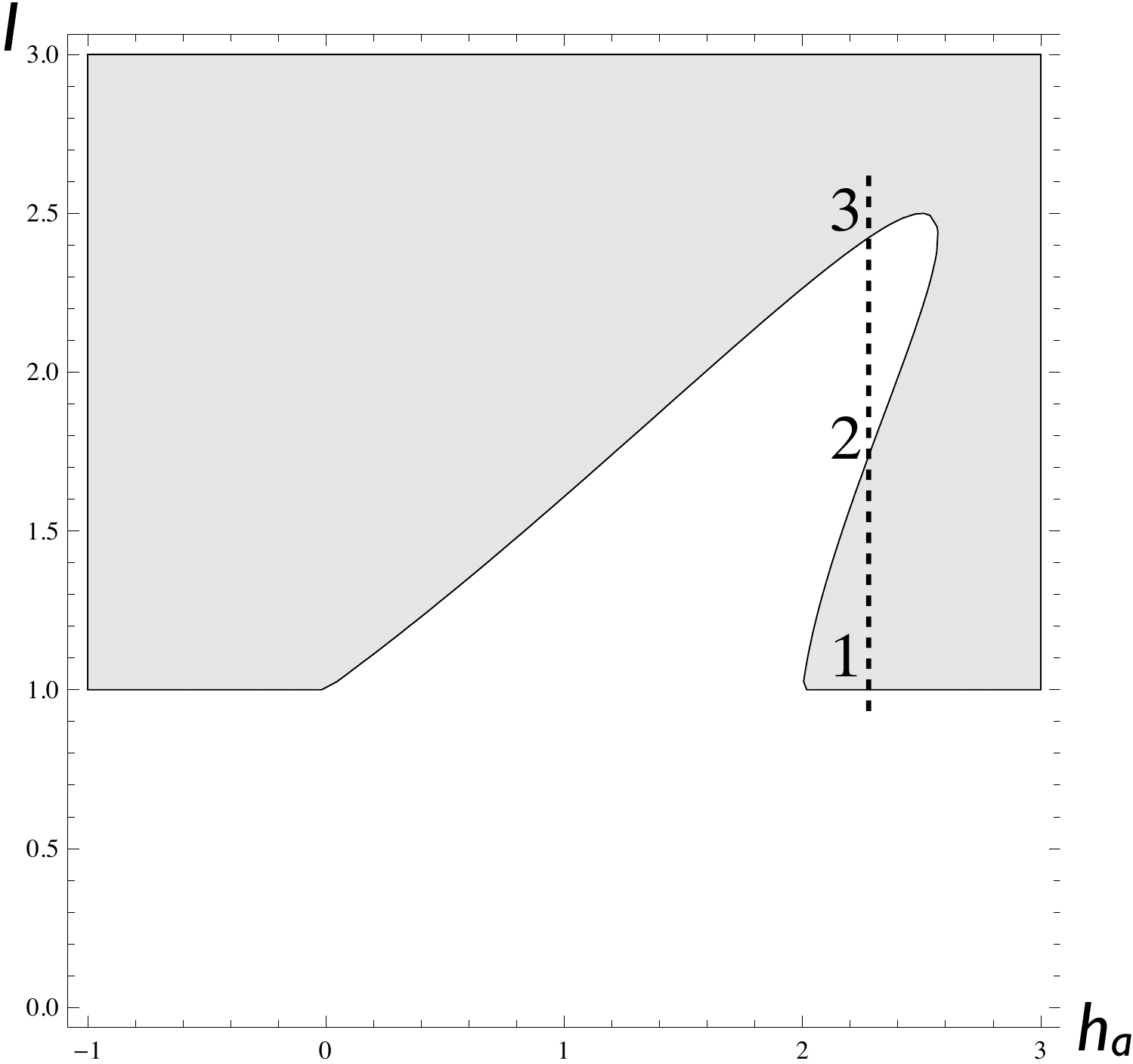}
\caption{Separation of the S state (white) and R state (grey) of the superconductor by a nonlinear function defined by $I'=1$.  The parameters of this system are $\mu=-1$, $\lambda=0.6$, $\nu,\kappa=0$, $K=\alpha=1$, and $\sigma=2$. The $1,2,3$ labels along the dashed line show the three places where the Josephson junction switches between superconducting and resistive states.}
\label{fig4}
\end{figure}

The salient differences between these fixed points and those found studying the fixed points of Eq.~(\ref{mzfDec}) are in the properties of $o_0$ and $o_\pi$. First, the transverse component of the ferromagnet is dynamic when $\mu\neq0$. Second, the static value of $\sin\phi$ is a nonlinear function of the current.  This results in a change in shape of the boundary separating the S and R states of the superconductor: See, for example, Fig.~\ref{fig4}, where the phase diagram develops a ``foldover region."  Consider the current increase at fixed magnetic field along the dashed line in Fig. \ref{fig4}.  The system undergoes changes from (1) S to R, (2) R to S, and (3) S to R again.  Unlike in a conventional Josephson junction, our model has multiple values of the current for which the junction changes between superconducting and resistive states.  Thus the junction has three `critical currents.'  Likewise at a particular fixed value of current, we can induce a change from R to S then S to R by increasing or decreasing the applied magnetic field.  This has no analogy in the RSJ model. As a function of the applied current, a rich variety of the coupled dynamics generally emerges, as seen in Fig.~\ref{fig5}, where we have plotted the stereographic projection of the magnetic direction.  A detailed analysis of this motion will be addressed in future work.  

\begin{figure}[pbh]
\includegraphics[width=0.49\linewidth,clip=]{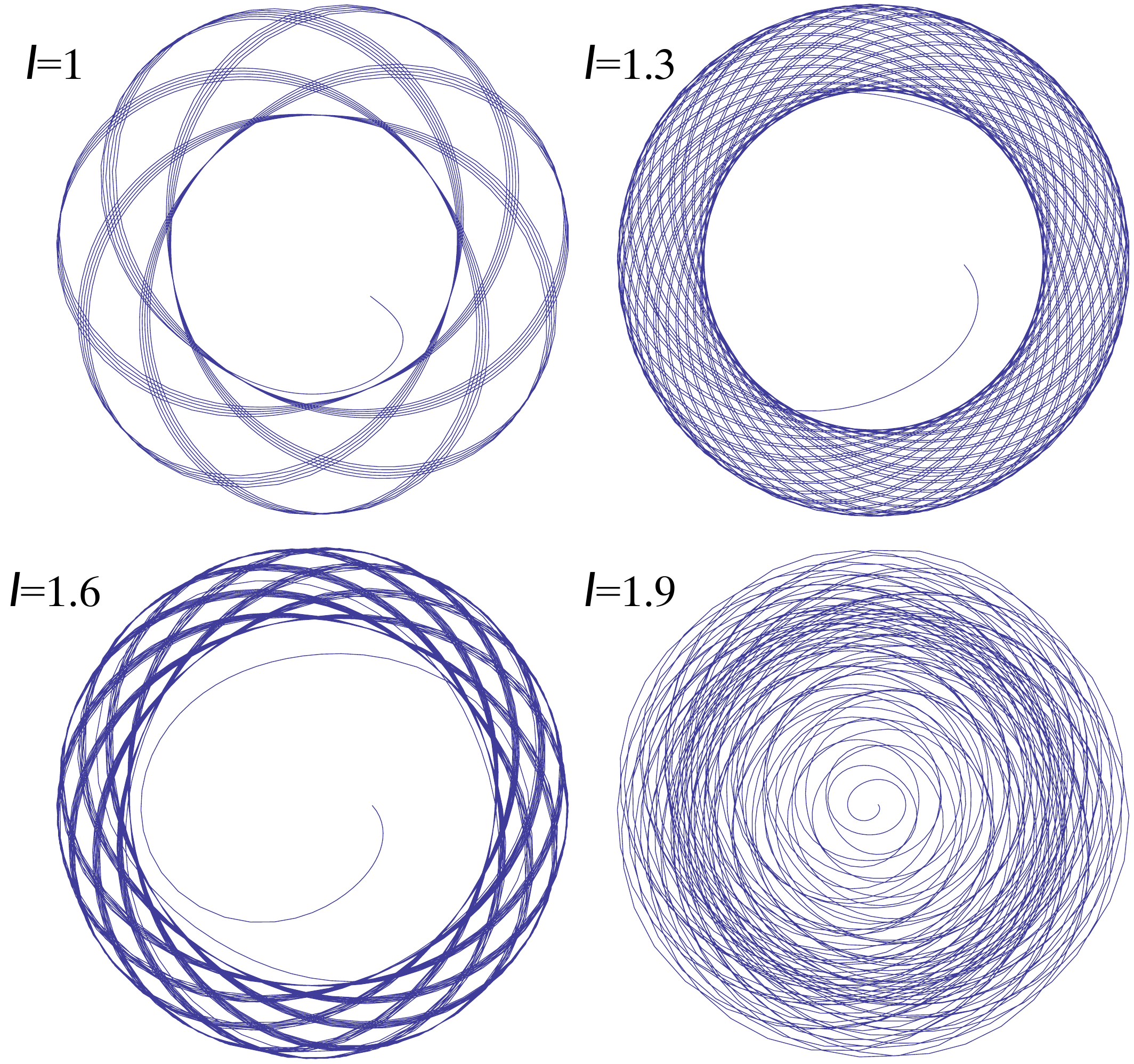}
\caption{Stereographic projection of the magnetization undergoing irreversible dynamics at different currents. Here, $\mu=0.1$, $\lambda=0.5$, $\nu,\kappa=0$, $K=1$, $\alpha=1$, and $\sigma=1$; initially positioned at $m_x=1$. Note that the scale is different between frames.}
\label{fig5}
\end{figure}

\end{widetext}

\end{document}